 \documentclass[preprint2]{aastex}

\newcommand{\myemail}{hstiele@mx.nthu.edu.tw}
\def\ie{i.\,e.}                                      
\def\eg{e.\,g.}                                      

\def\xmm{\textit{XMM-Newton}}
\def\4U1636{4U\,1636$-$53}

\shorttitle{milli-hertz variability in 4U\,1636--536}
\shortauthors{Stiele, Yu, Kong}

\begin{document}


\title{Millihertz Quasi-periodic Oscillations in 4U\,1636--536: Putting Possible Constraints on the Neutron Star Size}


\author{H. Stiele}
\affil{National Tsing Hua University, Department of Physics and Institute of Astronomy, No.~101 Sect.~2 Kuang-Fu Road,  30013, Hsinchu, Taiwan\\
Shanghai Astronomical Observatory and Center for Galaxy and Cosmology, Nandan Rod 80,  200030, Shanghai, China}
\email{\myemail}

\author{W. Yu}
\affil{Shanghai Astronomical Observatory and Center for Galaxy and Cosmology, Nandan Rod 80,  200030, Shanghai, China}
\email{wenfei@shao.ac.cn}

\and

\author{A. K. H. Kong}
\affil{National Tsing Hua University, Department of Physics and Institute of Astronomy, No.~101 Sect.~2 Kuang-Fu Road,  30013, Hsinchu, Taiwan}




\begin{abstract}
Based on previous studies of quasi-periodic oscillations in neutron star LMXBs, mHz quasi-periodic oscillations (QPO) are believed to be related to `marginally stable' burning on the neutron star (NS) surface. Our study of phase resolved energy spectra of these oscillations in \4U1636\ shows that the oscillations are not caused by variations in the blackbody temperature of the neutron star, but reveals a correlation between the change of the count rate during the mHz QPO pulse and the spatial extend of a region emitting blackbody emission. The maximum size of the emission area $R^2_{\mathrm{BB}}=216.7^{+93.2}_{-86.4}$km$^2$,  provides the direct evidence that the oscillations originate from a variable surface area constrained on the NS and are therefore not related to instabilities in the accretion disk. The obtained lower limit on the size of the neutron star (11.0 km) rules out equations of state that prefer small NS radii. Observations of mHz QPOs in NS LMXBs with NICER and eXTP will reduce the statistical uncertainty in the lower limit on the NS radius, which together with better estimates of the hardening factor and distance, will allow improving discrimination between different equations of state and compact star models. Furthermore, future missions will allow us to measure the peak blackbody emission area for a single mHz QPO pulse, which will push the lower limit to larger radii.
\end{abstract}



\section{Introduction}
In low-mass X-ray binaries (LMXBs) a variety of periodic and quasi-periodic phenomena has been observed \citep{2006csxs.book...39V,2014SSRv..183...43B}. This variability has mostly been associated with orbiting material in the accretion flow close to the compact object \citep[\eg][]{2011MNRAS.415.2323I}. In the case of a neutron star (NS) accretor, matter can accumulate on the NS surface and hence some of the variability phenomena can originate form the NS surface. Depending on the properties of the layer of accreted material (temperature, density, chemical composition of the accreted material) the ignition conditions for hydrogen or helium fusion can be reached and stable or unstable thermonuclear burning can proceed. These conditions are largely set by the mass accretion rate \citep{1981ApJ...247..267F,1998mfns.conf..419B}. If the accretion rate stays below a critical threshold, the accumulated material undergoes unstable nuclear burning, resulting in brief, intense thermonuclear X-ray bursts \citep[type-I bursts;][]{1993SSRv...62..223L,1995xrbi.nasa.....L,2006csxs.book..113S}. Theoretical models of hydrogen and helium burning suggest this threshold to be located around the Eddington limit, while observations place the transition of unstable to stable burning close to ten percent of the Eddington limit \citep{1981ApJ...247..267F,1988MNRAS.233..437V,1998mfns.conf..419B,2003A&A...405.1033C}. A possible solution for this discrepancy is to include mixing processes, \eg\ due to rotation and rotationally induced magnetic fields \citep{1993ApJ...419..768F,2004A&A...425..217Y,2007ApJ...663.1252P,2009A&A...502..871K}. Furthermore, the range of accretion rates in which the transition of unstable to stable burning takes place depends strongly on the composition of the burning layer and reaction rates \citep{2014ApJ...787..101K}.

At the transition between unstable and stable nuclear burning, the temperature dependence of the nuclear heating rate and cooling rate almost cancels. This leads to an oscillatory mode of burning called  `marginally stable burning'. Multi-zone numerical models of the NS envelope reproduce the oscillations at the transition from stable to unstable burning and derive the period of the oscillations to be the geometric mean of the accretion and thermal timescales for the burning layer, $\left(t_{\mathrm{them}}t_{\mathrm{accr}}\right)^{1/2}\sim100$s \citep{2007ApJ...665.1311H}.

A phenomenon speculated to be related to `marginally stable burning' on the NS surface are milli-hertz quasi-periodic oscillations (mHz QPOs) detected in NS LMXBs 
\citep{2001A&A...372..138R,2002ApJ...567L..67Y,2008ApJ...673L..35A,2012ApJ...748...82L}. Milli-hertz QPOs are observed at luminosities ($L_{2-20 keV}\simeq(5-11)\times10^{36}$\,erg\,s$^{-1}$) close to the transition luminosity between stable and unstable burning and their centroid frequencies are close to the oscillation periods found in models of `marginally stable burning'. 

Additional evidence that mHz QPOs are related to nuclear burning has been obtained from the observed anti-correlation between the kHz QPO frequency and count rate over one mHz QPO cycle, which suggests that the inner edge of the disk moves outward slightly as the luminosity increases during the cycle, perhaps consistent with enhanced radiation drag as the gas orbiting close to the NS is radiated by emission from the NS surface \citep{2002ApJ...567L..67Y}.  The connection between mHz QPOs and nuclear burning has been fostered by the frequency decrease of the mHz QPO in \4U1636\ observed prior to the occurrence of a type-I X-ray burst \citep{2008ApJ...673L..35A,2014MNRAS.445.3659L,2015MNRAS.454..541L}. This frequency drift can be reproduced by models of nuclear burning that take mixing processes and crustal heating into account, and could be caused by the cooling of deeper layers of the NS atmosphere \citep{2009A&A...502..871K}. 

\subsection{\4U1636}
\4U1636 is a persistent neutron star low-mass X-ray binary. It belongs to the class of atoll sources \citep{1989A&A...225...79H}. \4U1636\ shows intensity variations up to a factor of 10 on a $\sim$40~d cycle \citep{2005MNRAS.361..602S,2007MNRAS.379..247B}, following a narrow track in the color-color diagram and hardness-intensity diagram \citep{2007MNRAS.379..247B,2008ApJ...685..436A}. \4U1636\ was first observed as a strong continuous X-ray source with Copernicus \citep{1974MNRAS.169....7W} and Uhuru \citep{1974ApJS...27...37G}. Since then the source has been observed over a wide range of wavelengths. Optical photometry revealed an orbital period of $\sim$3.8~h \citep{1990A&A...234..181V}, and a companion star with a mass of $\sim$0.4~M$_{\odot}$, assuming an NS mass of 1.4~M$_{\odot}$ \citep{2002ApJ...568..279G}. X-ray studies of \4U1636\ have shown a variety of variability on different time scales ranging form mHz to kHz. The NS nature of the compact object has been confirmed through the detection of thermonuclear X-ray bursts \citep{1977ApJ...217L..23H,2006ApJ...639.1033G,2011MNRAS.413.1913Z}, which show asymptotic millisecond burst oscillations, indicating a spin frequency of $\sim$581~Hz \citep{1997IAUC.6541....1Z,2002ApJ...568..279G,2002ApJ...577..337S}. In addition to variability studies, the X-ray energy spectra, which show broad, asymmetric Fe emission lines, have been investigated \citep{2008ApJ...688.1288P,2010ApJ...720..205C,2010A&A...522A..96N,2013MNRAS.432.1144S}.

\begin{figure}
\centering
\resizebox{\hsize}{!}{\includegraphics[clip,angle=0]{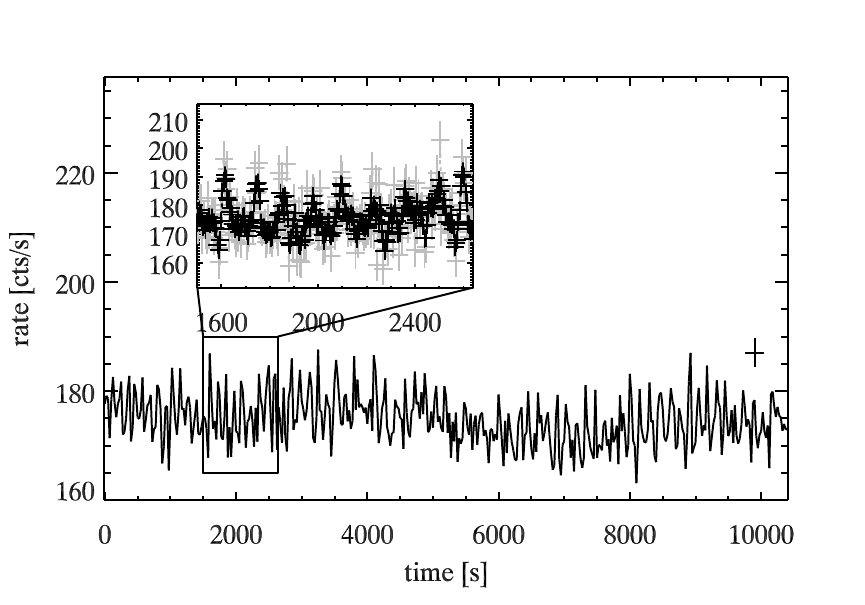}}
\caption{Light curve of the 10.4 ks of the September 2009 observation that show the mHz QPO. Each bin of the light curve contains 25~s. To enhance clarity of the plot error bars of individual bins are not shown, but the typical error bar is indicated. The inset shows the light curve with higher resolution: the grey points indicate bins of 5~s length, while the black points are smoothed by an additional factor of 3.}
\label{Fig:LC}
\end{figure}

\section{Observations \& Data analysis}
In this study we use two archival \xmm\ observations of \4U1636\ obtained on 2009  March 14 (Obs.ID: 0606070101) and September 5 (Obs.ID: 0606070301). The observations are taken with the EPIC/pn detector in timing mode, and have been included in the studies of \citet{2014MNRAS.445.3659L,2015MNRAS.454..541L}, who investigated the drift of the mHz QPO frequency, the evolution of spectral parameters with the drift of the frequency, and the duration of and spectral properties during the reappearance interval. We filter and extract the pn event file, using standard SAS (version 14.0.0) tools, paying particular attention to extract the list of photons not randomized in time. For our study we select the longest, continuous exposure available in each observation before a type-I X-ray burst occurred, which results in an exposure of 13.3 and 10.4 ks, respectively. We use the SAS task \texttt{epatplot} to investigate whether the observations are affected by pile-up. For the September observations we need to exclude the central three columns (37$\le$RAWX$\le$39) to minimize the effect of pile-up. To treat both observations in the same way and to obtain comparable (detector) count rates we exclude the same columns in the March observation. We extract powered density spectra (PDS) and light curves for each observation using columns 31$\le$RAWX$\le$36 and 40$\le$RAWX$\le$45. In the PDS, we verify that the noise level is consistent with the one expected for Poissonian noise \citep{1995ApJ...449..930Z} before we subtract the contribution due to Poissonian noise. Afterwards we normalize the PDS according to Leahy \citep{1983ApJ...272..256L} and convert it to square fractional rms \citep{1990A&A...227L..33B}. 

\begin{figure}
\centering
\resizebox{0.8\hsize}{!}{\includegraphics[clip,angle=0]{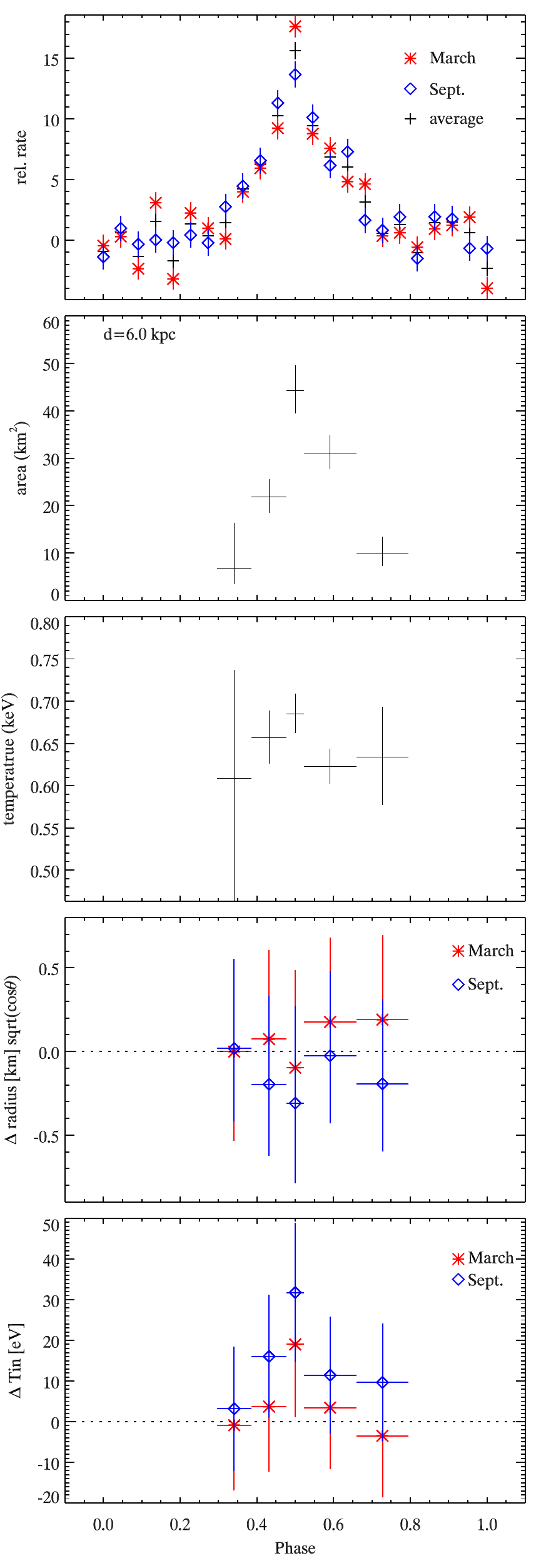}}
\caption{\footnotesize Upper panel: Waveform of the mHz QPO. To facilitate direct comparison of the waveform between both observations, the unpulsed level, obtained from $\phi\le0.3$ and $\phi\ge0.8$, have been subtracted. Middle two panels: Spectral parameters of the phase resolved spectra combining both observations and fitted with the ``quiescence" model and an additional bbodyrad model. Lower two panels: Spectral parameters of the fit of the phase resolved spectra with the  ``quiescence" model and allowing for a variable diskbb component. Given are difference between phase resolved and ``quiescence" inner disk radius and  temperature.}
\label{Fig:pulse_prof}
\end{figure}

\section{Light curve \& wave form}
From both observations we obtain light curves in which mHz QPOs, on a time scale of $\sim$140 s, are clearly present (see Fig.~\ref{Fig:LC}). In the September observation the mHz QPOs are observed before the occurrence of a type-I burst that is also covered by this observation \citep{2014MNRAS.445.3659L}. In addition, this observation covers about 20 ks after the type-I burst during which no mHz QPOs have been detected. Based on power density spectra we obtain an averaged centroid frequency in the range of 6.6 -- 7.2 mHz and 6.1 -- 6.5 mHz, respectively. 

We derive the mHz QPO waveform through a multi-step process \citep[similar to the approach used in][]{2002ApJ...567L..67Y}. At first we create a template of the waveform: We smooth the light curve by a factor of 3, \ie\ we replace the value in each bin by the averaged value obtained from this bin and its two adjacent neighbors. We then search for local maxima and minima in windows of 140 s of the smoothed light curve. We make sure that the extrema are alternating (\ie\ we excluded cases where there are \eg\ two maxima without a minimum in between), that the highest/lowest peak between two minima/maxima is picked as a maximum/minimum and that the maxima/minima exceed/fall below the averaged count rate by at least 30 per cent. We then aline the maxima found to obtain a first template of the waveform. To refine the waveform the template is correlated with the original (\ie\ not smoothed) light curve in steps of one bin width. A set of extrema is obtained using the best correlations in windows of 140 s and the maxima are alined to get a refined template of the waveform. The correlation and alining is repeated until a stable waveform is found. It takes 9(8) steps for the March(September) observation until the difference of the averaged count rate is less than 10$^{-3}$ cts/s in two consecutive steps. The obtained waveforms consist of a plateau of rather constant count rate, which we call the ``quiescence'' state of the mHz QPO, they then show a fast rise and a rather symmetric decline, followed by another plateau of rather constant count rate (Fig.~\ref{Fig:pulse_prof}). Both plateaus, at the beginning and end, occur at a similar count rate, and the pulse itself is located between $0.4\le\phi\le0.65$.  We want to point out that we assumed in this approach that the shape of the waveform stays the same despite the drift of the centroid frequency. To check if this assumption is reasonable, we divide the light curve in half and determine the waveform in each half. The waveform obtained from the half closer to the occurrence of the type-I burst agrees well with the waveform found in the other half. 

\section{Phase resolved spectra}
To obtain phase resolved energy spectra we use standard SAS tools and the start and end times of each bin from the light curves. During the extraction of energy spectra we pay special attention to generate ARF files of the pile-up corrected source region, following the steps laid down in the \xmm\ Users Guide\footnote{\url{http://xmm-tools.cosmos.esa.int/external/xmm\_user\_support/documentation/sas\_usg/USG/}}. To extract background spectra we use columns 3$\le$RAWX$\le$5. To get better statistics we regroup the bins (width:$0.0\bar{45}$) of the pulse into five bins, corresponding to early rise ($0.3<\phi<0.4$), rise ($0.4\le\phi<0.5$), peak ($\phi=0.5$), decay ($0.5<\phi\le0.65$) and late decay ($0.65<\phi<0.8$) of the pulse and extract an energy spectrum of each bin for both observations. We extract also a spectrum of the `quiescence' state of the mHz QPO, using $0.0\le\phi\le0.3$. The spectral fitting is done using \textsc{isis} V.~1.6.2 \citep{2000ASPC..216..591H}. We fit the `quiescence' spectra of both observations individually with an absorbed blackbody plus disk blackbody model, where the single blackbody accounts for the emission of the neutron star surface / boundary layer and the disk blackbody component for the emission of the accretion disk \citep{2013MNRAS.432.1144S}, to obtain a model of the X-ray emission of the accreting neutron star. We used the \texttt{tbabs} absorption model \citep{2000ApJ...542..914W} with abundances of \citet{1989GeCoA..53..197A} and the cross sections given in \citet{1992ApJ...400..699B}, with He cross-section based on \citet{1998ApJ...496.1044Y}. The obtained spectral parameters are given in Table~\ref{tab:specpar_bkg}. That the blackbody temperature does not show a correlation with the mHz QPO frequency has been reported \citep{2015MNRAS.454..541L}.

\begin{table*}[ht]
\caption{\label{tab:specpar_bkg}Spectral parameters of the background model: TBabs (bbodyrad+diskbb).}
\begin{center}
\begin{tabular}{lrr}
\hline\noalign{\smallskip}
\multicolumn{1}{c}{parameter} & \multicolumn{1}{c}{March} & \multicolumn{1}{c}{Sep.}   \\
 \hline\noalign{\smallskip}
N$_{H}$ [$10^{21}$ cm$^2$] &  $2.353\pm0.006$ &  $2.452\pm0.006$  \\
T$_{\mathrm{bbody}}$ [keV] &$ 1.81^{+0.03}_{-0.02} $& $1.75^{+0.02}_{-0.03}$ \\
A$_{\mathrm{bbody}}$ [km$^2$]$^{\dagger}$ & $10.9^{+0.7}_{-0.6} $ & $8.9\pm0.5$  \\
T$_{\mathrm{disk}}$ [keV] & $0.808\pm0.012$ & $0.772\pm0.011$\\
R$_{\mathrm{disk}}\times\sqrt{\cos{\theta}}$ [km]$^{\dagger}$  & $13.4\pm0.4$ & $12.5^{+0.4}_{-0.3}$\\
$\chi^2/\nu$ & $110.3/82$& $ 91.7/82$  \\
\noalign{\smallskip}
\hline
\end{tabular} 
\end{center}
Notes: $^{\dagger}$: derived from normalization assuming a distance of 6.0 kpc \citep{2006ApJ...639.1033G} and a mass of 1.4 M$_{\odot}$. 
\end{table*}

\begin{figure}
\centering
\resizebox{\hsize}{!}{\includegraphics[clip,angle=0]{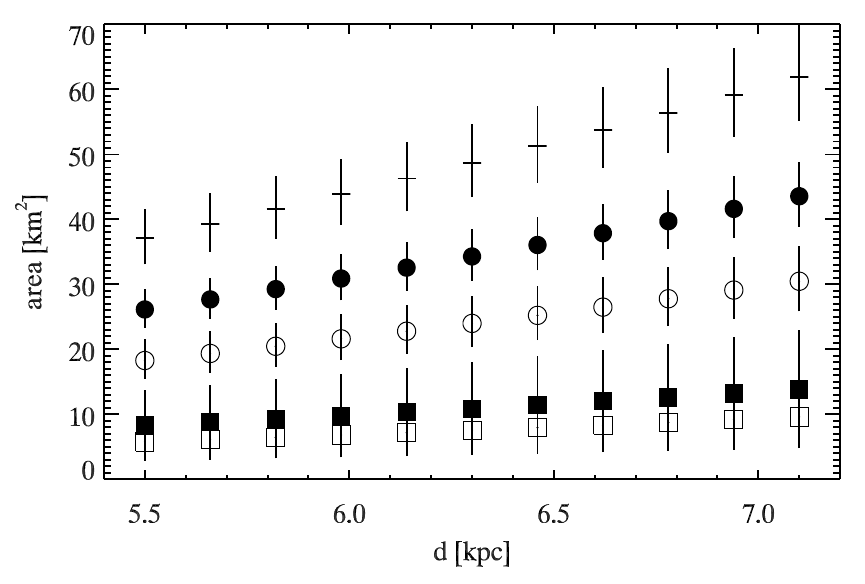}}
\caption{Dependence of the absolute value of the apparent  blackbody emission area on the distance to the source. We used a distance of $6.0^{+1.1}_{-0.5}$ kpc \citep{2006ApJ...639.1033G}, where the upper error is bigger than the lower error as it contains additional uncertainty on the mass of the neutron star. Different symbols denote different phases as follows: open squares: $0.3\le\phi<0.4$, open circles: $0.4\le\phi<0.5$, crosses:  $\phi=0.5$, filled circles: $0.5<\phi\le0.65$, filled squares: $0.65\le\phi<0.8$.}
\label{Fig:distdep}
\end{figure}

In \citet{2007ApJ...665.1311H} the mHz QPO is interpreted as an oscillatory burning mode across the whole neutron star surface. According to this interpretation the oscillations come from variations in the blackbody temperature. To test this picture we fit the phase resolved energy spectra with the model of the ``quiescence'' spectra, leaving the blackbody temperature free. We find a change of the blackbody temperature (on the 10 eV scale) with the pulse profile, but the obtained reduced $\chi^2$ values are unacceptable large (Tabel \ref{tab:specpar}) and the spectral residuals clearly show that an additional component at lower energies is present.

Next we try to fit the phase resolved energy spectra with the obtained ``quiescence'' spectra and an additional single blackbody model (\texttt{bbodyrad}). We fit the energy spectra of both observations individually. In addition, we perform joint fits of energy spectra belonging to the same phase bin of both observations simultaneously. In the case of fitting both observations simultaneously, we use the individual ``quiescence" model of each observation and assume that the parameters of the added blackbody model are the same for both observations. Already from the spectral fits of the individual observations it is evident that the emission region of the black body emission changes during the pulse, while the temperature remains constant within error bars. To increase statistics we combined data of both observations and obtained a constant temperature of 0.65$\pm$0.05 keV\footnote{We note here that \citet{2001A&A...372..138R} found that in 4U1608--52 the ``variable'' spectral component, which is related to the mHz QPO, is much softer than the averaged spectrum.} while the apparent area increased from $6.8^{+9.5}_{-3.4}$ km$^2$ at the begin and $9.9^{+3.6}_{-2.7}$ km$^2$ at the end of the pulse to $44.2^{+5.4}_{-4.8}$ km$^2$ during the pulse (Fig.~\ref{Fig:pulse_prof}), assuming a distance to the source of 6.0 kpc \citep{2006ApJ...639.1033G}, with $\chi^2_{\mathrm{red}}$ between 1.0 and 1.3. The spectral fitting results for each bin can be found in Table~\ref{tab:specpar}. For the March observation we cannot obtain constrained spectral parameters for the early rise ($0.3<\phi<0.4$). 

How the uncertainty in the distance affects the size of the apparent area is shown in Fig.~\ref{Fig:distdep}. As the distance remains the same during the mHz QPO cycle, uncertainties in the distance affect the absolute value of the apparent area, but not the evolution of the apparent area with mHz QPO phase. The apparent area ($R^2_{\infty}$) is related to the area measured at the stellar surface ($R^2_{\mathrm{BB}}$) by the following relation \citep{1985ApJ...299..487S}: \[R^2_{\mathrm{BB}} = R^2_{\infty} \times f_{\mathrm{col}}^4 \times \left(1-2\frac{M}{R_{\mathrm{NS}}}\right)\] where $G=c=1$. Based on the narrow luminosity range in which mHz QPOs are observed and assuming a He enriched environment \citep{2015MNRAS.454..541L} we assume a hardening factor $f_{\mathrm{col}}$ of $1.60^{+0.10}_{-0.15}$. The value has been obtained by fitting the atmosphere spectra for different effective gravity values and compositions in the $0.5-1.1\times10^{37}$\,erg\,s$^{-1}$ luminosity range from \citet{2012A&A...545A.120S} with a diluted blackbody in the 1 -- 10 keV range, appropriate for the EPIC/pn instrument. The compactness of the neutron star has been derived as  $M/R_{\mathrm{NS}}=\beta_{\mathrm{avg}}=0.126$ with an upper limit of $0.163$ by \citet{2002ApJ...564..353N} based on RXTE observations of the bolometric flux oscillations that occur during the rise of X-ray bursts. They assumed two antipodal hot spots that expand linearly with time. Models with just one hot spot did no allow to constrain the compactness. Relativistic Doppler shifts and aberration due to rotational motion of the hot spots have been neglected. They also assumed that the background was known. Using the above values, the area covered on the neutron star surface during pulse peak corresponds to $R^2_{\mathrm{BB}}=216.7^{+93.2}_{-86.4}$km$^2$. This finding allows to determine a rough lower limit on the neutron star radius of $14.7^{+2.9}_{-3.3}$km, not taking into account additional geometrical effects \citep[see \eg\ ][]{2000ApJ...544..390P}, anisotropy \citep{2016ApJ...819...47H,2016ApJ...818..135C} and uncertainties on the distance, and shows that the process that causes the variability seen as mHz QPO originates in spatial expansion of the zone of nuclear burning on the neutron star surface.  
Relating the change of the emission area to the rotation of the neutron star can be excluded as the neutron star has a spin frequency of $\sim$518~Hz \citep{2002ApJ...577..337S}.

In an alternative scenario, the origin of the mHz QPO is related to thermal/viscous instabilities in the accretion disk. In GRS\,1915+105 variability on time scales of 100 -- 1000 s has been observed that has been attributed to disk instabilities \citep{1997ApJ...485L..83T}. However, the shape of the light curve, a gradual increase through a shoulder followed by a sharp drop and a secondary and in some cases even a tertiary peak observed in GRS\,1919+105 differs significantly from the rather symmetric profile we find in the case of 4U\,1636--53. We also fit the phase resolved energy spectra with the ``quiescence'' model, keeping the absorption and the parameters of the blackbody emission fixed at the values found during the ``quiescence'' state of the mHz QPO. The parameters of the disk blackbody model are allowed to vary to investigate whether changes in the spectra can be attributed to the emission of the disk component. The inner disk radius stays constant over the mHz QPO cycle (Fig.~\ref{Fig:pulse_prof} and Table~\ref{tab:specpar}). For the inner disk temperature we find an increase of about 20--30 eV during the central bin, which corresponds to the peak of the mHz QPO, while for all other bins the temperature agrees with the one of the ``quiescence'' spectrum. Taking uncertainties on the temperatures into account, the increase during the central peak is significant on a level of 1.06$\sigma$ in the March and of 1.85$\sigma$ in the September observation. In GRS\,1915+105 temperature changes between the shoulder and the peaks have been found that are more than a factor ten larger than the one seen here \citep{1997ApJ...485L..83T}. Taking the constant inner disk radius, the negligible change in inner disk temperature and a maximum extension of the blackbody emission area consistent with the size of a NS an origin of the mHz QPO related to instabilities in the accretion disk is highly unlikely.

We would like to mention that for the March data that are not affected by pile-up we also extract light curves and power density and energy spectra form the whole 31$\le$RAWX$\le$45 range, \ie\ without excluding the central three columns. Including the central three columns increases the amount of photons and hence the count rate. The shapes of the waveform and the phase resolved energy spectra are consistent with the one obtained from the data where the central three columns are excluded and therefore the results presented here are neither affected by the pile-up correction, nor by pile-up itself.

\begin{table*}[ht]
\caption{\label{tab:specpar}Spectral parameters used in fitting the phase resolved spectra.}
\begin{center}
\footnotesize
\begin{tabular}{lrrrrr}
\tableline\noalign{\smallskip}
\multicolumn{1}{c}{parameter} & \multicolumn{1}{c}{$0.3\le\phi<0.4$} & \multicolumn{1}{c}{$0.4\le\phi<0.5$} & \multicolumn{1}{c}{$\phi=0.5$}  &  \multicolumn{1}{c}{$0.5<\phi\le0.65$} & \multicolumn{1}{c}{$0.65\le\phi<0.8$}  \\
 \tableline\noalign{\smallskip}
\multicolumn{6}{c}{both observations; additional bbodyrad model} \\
 \tableline\noalign{\smallskip}
temperature [keV] & $0.61_{-0.15}^{+0.13}$ & $0.66\pm0.03$ & $0.69\pm0.02$ & $0.62\pm0.02$& $0.63\pm0.06$ \\
\smallskip
area [km$^2$]$^{\dagger}$& $6.9_{-3.5}^{+9.5}$& $21.8_{-3.3}^{+3.8}$ & $44.2_{-4.8}^{+5.4}$ & $31.1_{-3.3}^{+3.8}$& $9.9^{+3.6}_{-2.7}$   \\
\smallskip
$\chi^2/\nu$ & $227.2/172$ & $211.9/172$ & $174.2/172$ & $224.9/172$ & $219.1/172$ \\
\tableline\noalign{\smallskip}
\multicolumn{6}{c}{March; additional bbodyrad model}  \\
\noalign{\smallskip}
\tableline\noalign{\smallskip}
temperature [keV] &  $-$ &  $0.62^{+0.06}_{-0.05}$ & $0.67^{+0.04}_{-0.03}$ & $0.61\pm0.03$&  $0.54^{+0.07}_{-0.06}$  \\
\smallskip
area [km$^2$]$^{\dagger}$&  $-$&  $25.0^{+8.7}_{-6.5}$ & $48.4^{+9.2}_{-7.8}$ & $38.0^{+6.8}_{-5.8}$ &  $19.9^{+10.4}_{-6.8}$ \\
\smallskip
$\chi^2/\nu$ & $-$ & $95.5/85$ & $111.7/85$ & $123.6/85$ & $98.3/85$ \\
\tableline\noalign{\smallskip}
\multicolumn{6}{c}{September; additional bbodyrad model}  \\
\noalign{\smallskip}
\tableline\noalign{\smallskip}
temperature [keV] &  $0.62^{+0.11}_{-0.12}$& $0.68\pm0.04$ & $0.70\pm0.03$ & $0.64\pm0.03$ &  $0.76^{+0.10}_{-0.09}$\\
\smallskip
area [km$^2$]$^{\dagger}$&  $10.1^{+10.4}_{-4.5}$& $20.7_{-3.7}^{+4.4}$ & $41.4^{+6.7}_{-5.8}$ & $26.7^{+4.5}_{-3.9}$ &  $5.2^{+2.7}_{-1.8}$ \\
\smallskip
$\chi^2/\nu$ & $111.5/85$& $115.7/85$ & $64.3/85$ & $95.8/85$ & $115.5/85$ \\
\noalign{\smallskip}
\tableline
 \tableline\noalign{\smallskip}
\multicolumn{6}{c}{March; bbodyrad model}  \\
\noalign{\smallskip}
\tableline\noalign{\smallskip}
temperature [keV] &  $1.814\pm0.005$ &  $1.826\pm0.004$ & $1.853^{+0.006}_{-0.007}$ & $1.823\pm0.004$&  $1.809\pm0.004$  \\
\smallskip
$\chi^2/\nu$ & $105.7/86$ & $179.0/86$ & $317.0/86$ & $403.0/86$ & $135.4/86$ \\
\tableline\noalign{\smallskip}
\multicolumn{6}{c}{September; bbodyrad model}  \\
\noalign{\smallskip}
\tableline\noalign{\smallskip}
temperature [keV] &  $1.749\pm0.004 $& $1.763\pm0.004$ & $1.796\pm0.006$ & $1.767\pm0.003$ &  $1.753\pm0.004$\\
\smallskip
$\chi^2/\nu$ & $137.0/86$& $275.6/86$ & $378.2/86$ & $373.4/86$ & $150.4/86$ \\
\noalign{\smallskip}
\tableline
 \tableline\noalign{\smallskip}
\multicolumn{6}{c}{March; diskbb model}  \\
\noalign{\smallskip}
\tableline\noalign{\smallskip}
T$_{\mathrm{disk}}$ [keV] & $0.807\pm0.004$ & $0.812\pm0.004$ & $0.827\pm0.006$ & $0.811\pm0.003$ & $0.804\pm0.003$ \\
\smallskip
R$_{\mathrm{disk}}\times\sqrt{\cos{\theta}}$ [km]$^{\dagger}$& $13.4\pm0.1$ & $13.5\pm0.1$ & $13.3\pm0.2$ & $13.6\pm0.1$ & $13.6\pm0.1$  \\
\smallskip
$\chi^2/\nu$ & 105.9/85 & 95.6/85 & 111.3/85 & 125.0/85 & 98.2/85\\
\smallskip
$\Delta$T$_{\mathrm{disk}}$ [keV] & $-0.001\pm0.016$ & $0.004\pm0.016$ & $0.019\pm0.018$ & $0.003\pm0.015$ & $-0.004\pm0.015$ \\
\smallskip
$\Delta$R$_{\mathrm{disk}}\times\sqrt{\cos{\theta}}$ [km]$^{\dagger}$& $0.0\pm0.5$ & $0.1\pm0.5$ & $-0.1\pm0.6$ & $0.2\pm0.5$ & $0.2\pm0.5$  \\
\tableline\noalign{\smallskip}
\multicolumn{6}{c}{September; diskbb model}  \\
\noalign{\smallskip}
\tableline\noalign{\smallskip}
T$_{\mathrm{disk}}$ [keV] & $0.775\pm0.004$ & $0.787\pm0.004$ & $0.804\pm0.006$ & $0.783\pm0.003$ & $0.782\pm0.003$\\
\smallskip
R$_{\mathrm{disk}}\times\sqrt{\cos{\theta}}$ [km]$^{\dagger}$& $12.5\pm0.1$  & $12.3\pm0.1$  & $12.2\pm0.2$ & $12.5\pm0.1$  & $12.3\pm0.1$  \\
\smallskip
$\chi^2/\nu$ & 112.0/85 & 118.0/85 & 64.1/85 & 97.3/85 & 115.8/85\\
\smallskip
$\Delta$T$_{\mathrm{disk}}$ [keV] & $0.003\pm0.015$ & $0.015\pm0.015$ & $0.032\pm0.017$ & $0.011\pm0.014$ & $0.010\pm0.014$\\
\smallskip
$\Delta$R$_{\mathrm{disk}}\times\sqrt{\cos{\theta}}$ [km]$^{\dagger}$& $0.0^{+0.5}_{-0.4}$  & $-0.2^{+0.5}_{-0.4}$  & $-0.3^{+0.6}_{-0.5}$ & $0.0^{+0.5}_{-0.4}$  & $-0.2^{+0.5}_{-0.4}$  \\
\noalign{\smallskip}
\tableline
\end{tabular} 
\end{center}
{Notes: $^{\dagger}$: derived from normalization assuming a distance of 6.0 kpc \citep{2006ApJ...639.1033G} and a mass of 1.4 M$_{\odot}$.}
\end{table*}

\begin{figure}
\centering
\resizebox{\hsize}{!}{\includegraphics[clip,angle=0,scale=0.8]{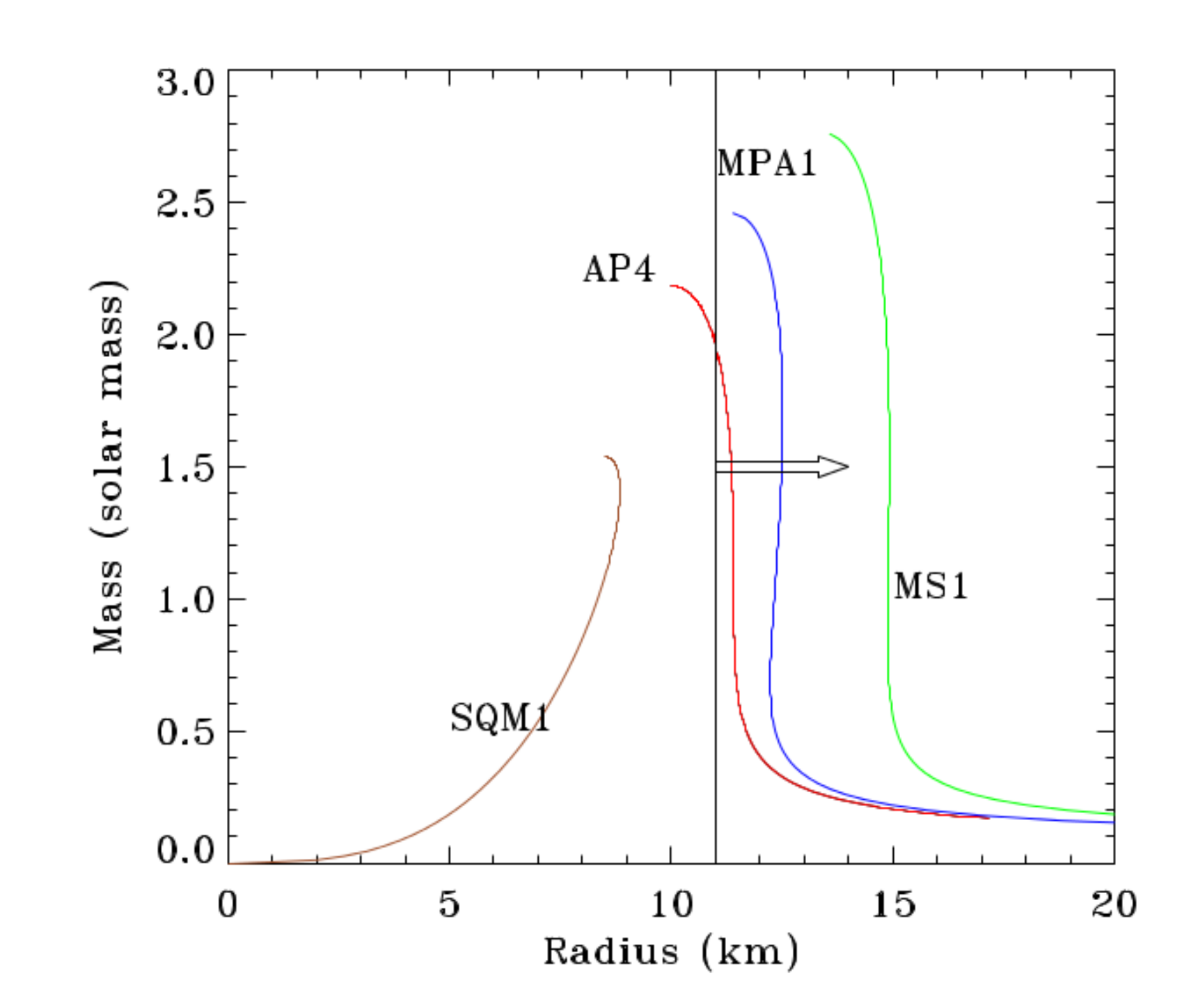}}
\caption{Mass-radius curves for different equation of states \citep{2016arXiv160302698O}. The lower limit on the NS radius obtained based on current estimates of the source distance allows us to rule out NS EoS that prefer a small NS radius.}
\label{Fig:EoS}
\end{figure}

\section{Constraints on neutron star size and equation of state}
The fact that the emission of the mHz QPOs can be described by a blackbody model shows that mHz QPOs are related to some process that produces thermal emission. The correlation between the size of the emission region of the blackbody emission and the mHz QPO phase provides evidence that mHz QPOs do not origin in temperature oscillations of the whole neutron star, but are related to nuclear burning on the NS surface. The maximum size of the blackbody emission found during the mHz QPO peak shows that the emission in a mHz QPO originates only in parts of the NS surface. Putting the results of our phase resolved spectral study together with what we already know about mHz QPOs from previous studies and including our knowledge about type-I X-ray bursts we arrive at the following picture. During the accretion of matter form the companion star onto the NS surface the conditions to ignite `marginally stable' nuclear burning are reached locally. As a result we detect spatially confined bursts of thermal emission as mHz QPOs. One way to get confinement of burning are magnetic fields \citep{1998ApJ...496..915B}. When the NS has accreted enough matter, `marginally stable'  burring is suppressed and an explosive nuclear burning spreading over the whole NS surface takes place, which is observed as a type-I X-ray burst. After the type-I burst all the matter to fuel nuclear burning is exhausted and no mHz QPOs are detected until enough matter to start `marginally stable'  burring has been accreted again.

The lower limit on the NS radius (11.0 km; 2$\sigma$ lower limit) we got based on current estimates of the source distance allows us to rule out NS equation of states (EoS) that prefer a small NS radius, \eg\ strange quark matter \citep[Fig~\ref{Fig:EoS}; see][for a recent review of NS EoS]{2016arXiv160302698O}. This lower limit includes the uncertainties on the apparent size of the emission area, the hardening factor, and the compactness of the NS. Including the uncertainty on the distance estimation, the lower limit on the NS radius remains above 10 km.
Observations of one life cycle of the mHz QPO in \4U1636\ \citep[19 ks;][]{2015MNRAS.454..541L} with NICER will allow to reduce the statistical uncertainties in the radius to about 0.3 km. With the Low energy Focusing Array (LFA) onboard the enhanced X-ray Timing and Polarisation mission (eXTP) it will be possible to reduce the uncertainties even further to $\la0.15$ km, and to constrain the NS radius on $\pm1$ km using current estimates of the source distance and hardening factor. Given improved constrains on the hardening factor (which will be also obtained with NICER and eXTP) and on the distance, observations of mHz QPOs in NS LMXBs with eXTP and NICER will provide solid lower limits on the NS radius and improve discrimination between different EoS and between neutron star and quark star models. Future missions will also allow us to measure the emission area for single mHz QPO pulses. The maximum peak blackbody area of a single mHz QPO pulse will push the lower limit to larger radii than the current results based on an averaged mHz QPO waveform.

\acknowledgments

This work is based on observations obtained with \xmm, an ESA science mission with instruments and contributions directly funded by ESA Member States and NASA. Data used in this work are available from the \xmm\ Science Archive (http://www.cosmos.esa.int/web/xmm-newton/xsa). This work was supported by the National Natural Science Foundation of China under grant No. 11073043, 11333005, and 11350110498, by Strategic Priority Research Program "The Emergence of Cosmological Structures" under Grant No. XDB09000000 and the XTP project under Grant No. XDA04060604, by the Shanghai Astronomical Observatory Key Project and by the Chinese Academy of Sciences Fellowship for Young International Scientists Grant. This project was supported by the Ministry of Science and Technology of the Republic of China (Taiwan) through grants 103-2628-M-007-003-MY3 and 104-281-M-007-060.

{\it Facilities:} \facility{\xmm}.


\begin{thebibliography}{}
\expandafter\ifx\csname natexlab\endcsname\relax\def\natexlab#1{#1}\fi

\bibitem[{{Altamirano} {et~al.}(2008{\natexlab{a}}){Altamirano}, {van der
  Klis}, {M{\'e}ndez}, {Jonker}, {Klein-Wolt}, \&
  {Lewin}}]{2008ApJ...685..436A}
{Altamirano}, D., {van der Klis}, M., {M{\'e}ndez}, M., {et~al.}
  2008{\natexlab{a}}, \apj, 685, 436

\bibitem[{{Altamirano} {et~al.}(2008{\natexlab{b}}){Altamirano}, {van der
  Klis}, {Wijnands}, \& {Cumming}}]{2008ApJ...673L..35A}
{Altamirano}, D., {van der Klis}, M., {Wijnands}, R., \& {Cumming}, A.
  2008{\natexlab{b}}, \apjl, 673, L35

\bibitem[{{Anders} \& {Grevesse}(1989)}]{1989GeCoA..53..197A}
{Anders}, E., \& {Grevesse}, N. 1989, \gca, 53, 197

\bibitem[{{Balucinska-Church} \& {McCammon}(1992)}]{1992ApJ...400..699B}
{Balucinska-Church}, M., \& {McCammon}, D. 1992, \apj, 400, 699

\bibitem[{{Belloni} \& {Hasinger}(1990)}]{1990A&A...227L..33B}
{Belloni}, T., \& {Hasinger}, G. 1990, \aap, 227, L33

\bibitem[{{Belloni} {et~al.}(2007){Belloni}, {Homan}, {Motta}, {Ratti}, \&
  {M{\'e}ndez}}]{2007MNRAS.379..247B}
{Belloni}, T., {Homan}, J., {Motta}, S., {Ratti}, E., \& {M{\'e}ndez}, M. 2007,
  \mnras, 379, 247

\bibitem[{{Belloni} \& {Stella}(2014)}]{2014SSRv..183...43B}
{Belloni}, T.~M., \& {Stella}, L. 2014, \ssr, 183, 43

\bibitem[{{Bildsten}(1998)}]{1998mfns.conf..419B}
{Bildsten}, L. 1998, in NATO ASIC Proc. 515: The Many Faces of Neutron Stars.,
  ed. R.~{Buccheri}, J.~{van Paradijs}, \& A.~{Alpar}, 419

\bibitem[{{Brown} \& {Bildsten}(1998)}]{1998ApJ...496..915B}
{Brown}, E.~F., \& {Bildsten}, L. 1998, \apj, 496, 915

\bibitem[{{Cackett} {et~al.}(2010){Cackett}, {Miller}, {Ballantyne}, {Barret},
  {Bhattacharyya}, {Boutelier}, {Miller}, {Strohmayer}, \&
  {Wijnands}}]{2010ApJ...720..205C}
{Cackett}, E.~M., {Miller}, J.~M., {Ballantyne}, D.~R., {et~al.} 2010, \apj,
  720, 205

\bibitem[{{Chenevez} {et~al.}(2016){Chenevez}, {Galloway}, {in 't Zand},
  {Tomsick}, {Barret}, {Chakrabarty}, {F{\"u}rst}, {Boggs}, {Christensen},
  {Craig}, {Hailey}, {Harrison}, {Romano}, {Stern}, \&
  {Zhang}}]{2016ApJ...818..135C}
{Chenevez}, J., {Galloway}, D.~K., {in 't Zand}, J.~J.~M., {et~al.} 2016, \apj,
  818, 135

\bibitem[{{Cornelisse} {et~al.}(2003){Cornelisse}, {in't Zand}, {Verbunt},
  {Kuulkers}, {Heise}, {den Hartog}, {Cocchi}, {Natalucci}, {Bazzano}, \&
  {Ubertini}}]{2003A&A...405.1033C}
{Cornelisse}, R., {in't Zand}, J.~J.~M., {Verbunt}, F., {et~al.} 2003, \aap,
  405, 1033

\bibitem[{{Fujimoto}(1993)}]{1993ApJ...419..768F}
{Fujimoto}, M.~Y. 1993, \apj, 419, 768

\bibitem[{{Fujimoto} {et~al.}(1981){Fujimoto}, {Hanawa}, \&
  {Miyaji}}]{1981ApJ...247..267F}
{Fujimoto}, M.~Y., {Hanawa}, T., \& {Miyaji}, S. 1981, \apj, 247, 267

\bibitem[{{Galloway} {et~al.}(2006){Galloway}, {Psaltis}, {Muno}, \&
  {Chakrabarty}}]{2006ApJ...639.1033G}
{Galloway}, D.~K., {Psaltis}, D., {Muno}, M.~P., \& {Chakrabarty}, D. 2006,
  \apj, 639, 1033

\bibitem[{{Giacconi} {et~al.}(1974){Giacconi}, {Murray}, {Gursky}, {Kellogg},
  {Schreier}, {Matilsky}, {Koch}, \& {Tananbaum}}]{1974ApJS...27...37G}
{Giacconi}, R., {Murray}, S., {Gursky}, H., {et~al.} 1974, \apjs, 27, 37

\bibitem[{{Giles} {et~al.}(2002){Giles}, {Hill}, {Strohmayer}, \&
  {Cummings}}]{2002ApJ...568..279G}
{Giles}, A.~B., {Hill}, K.~M., {Strohmayer}, T.~E., \& {Cummings}, N. 2002,
  \apj, 568, 279

\bibitem[{{Hasinger} \& {van der Klis}(1989)}]{1989A&A...225...79H}
{Hasinger}, G., \& {van der Klis}, M. 1989, \aap, 225, 79

\bibitem[{{He} \& {Keek}(2016)}]{2016ApJ...819...47H}
{He}, C.-C., \& {Keek}, L. 2016, \apj, 819, 47

\bibitem[{{Heger} {et~al.}(2007){Heger}, {Cumming}, \&
  {Woosley}}]{2007ApJ...665.1311H}
{Heger}, A., {Cumming}, A., \& {Woosley}, S.~E. 2007, \apj, 665, 1311

\bibitem[{{Hoffman} {et~al.}(1977){Hoffman}, {Lewin}, \&
  {Doty}}]{1977ApJ...217L..23H}
{Hoffman}, J.~A., {Lewin}, W.~H.~G., \& {Doty}, J. 1977, \apjl, 217, L23

\bibitem[{{Houck} \& {Denicola}(2000)}]{2000ASPC..216..591H}
{Houck}, J.~C., \& {Denicola}, L.~A. 2000, in Astronomical Society of the
  Pacific Conference Series, Vol. 216, Astronomical Data Analysis Software and
  Systems IX, ed. {N.~Manset, C.~Veillet, \& D.~Crabtree}, 591

\bibitem[{{Ingram} \& {Done}(2011)}]{2011MNRAS.415.2323I}
{Ingram}, A., \& {Done}, C. 2011, \mnras, 415, 2323

\bibitem[{{Keek} {et~al.}(2014){Keek}, {Cyburt}, \&
  {Heger}}]{2014ApJ...787..101K}
{Keek}, L., {Cyburt}, R.~H., \& {Heger}, A. 2014, \apj, 787, 101

\bibitem[{{Keek} {et~al.}(2009){Keek}, {Langer}, \& {in't
  Zand}}]{2009A&A...502..871K}
{Keek}, L., {Langer}, N., \& {in't Zand}, J.~J.~M. 2009, \aap, 502, 871

\bibitem[{{Leahy} {et~al.}(1983){Leahy}, {Elsner}, \&
  {Weisskopf}}]{1983ApJ...272..256L}
{Leahy}, D.~A., {Elsner}, R.~F., \& {Weisskopf}, M.~C. 1983, \apj, 272, 256

\bibitem[{{Lewin} {et~al.}(1993){Lewin}, {van Paradijs}, \&
  {Taam}}]{1993SSRv...62..223L}
{Lewin}, W.~H.~G., {van Paradijs}, J., \& {Taam}, R.~E. 1993, \ssr, 62, 223

\bibitem[{{Lewin} {et~al.}(1995){Lewin}, {van Paradijs}, \& {van den
  Heuvel}}]{1995xrbi.nasa.....L}
{Lewin}, W.~H.~G., {van Paradijs}, J., \& {van den Heuvel}, E.~P.~J. 1995,
  X-ray Binaries

\bibitem[{{Linares} {et~al.}(2012){Linares}, {Altamirano}, {Chakrabarty},
  {Cumming}, \& {Keek}}]{2012ApJ...748...82L}
{Linares}, M., {Altamirano}, D., {Chakrabarty}, D., {Cumming}, A., \& {Keek},
  L. 2012, \apj, 748, 82

\bibitem[{{Lyu} {et~al.}(2014){Lyu}, {M{\'e}ndez}, \&
  {Altamirano}}]{2014MNRAS.445.3659L}
{Lyu}, M., {M{\'e}ndez}, M., \& {Altamirano}, D. 2014, \mnras, 445, 3659

\bibitem[{{Lyu} {et~al.}(2015){Lyu}, {M{\'e}ndez}, {Zhang}, \&
  {Keek}}]{2015MNRAS.454..541L}
{Lyu}, M., {M{\'e}ndez}, M., {Zhang}, G., \& {Keek}, L. 2015, \mnras, 454, 541

\bibitem[{{Nath} {et~al.}(2002){Nath}, {Strohmayer}, \&
  {Swank}}]{2002ApJ...564..353N}
{Nath}, N.~R., {Strohmayer}, T.~E., \& {Swank}, J.~H. 2002, \apj, 564, 353

\bibitem[{{Ng} {et~al.}(2010){Ng}, {D{\'{\i}}az Trigo}, {Cadolle Bel}, \&
  {Migliari}}]{2010A&A...522A..96N}
{Ng}, C., {D{\'{\i}}az Trigo}, M., {Cadolle Bel}, M., \& {Migliari}, S. 2010,
  \aap, 522, A96

\bibitem[{{Ozel} \& {Freire}(2016)}]{2016arXiv160302698O}
{Ozel}, F., \& {Freire}, P. 2016, ArXiv e-prints, arXiv:1603.02698

\bibitem[{{Pandel} {et~al.}(2008){Pandel}, {Kaaret}, \&
  {Corbel}}]{2008ApJ...688.1288P}
{Pandel}, D., {Kaaret}, P., \& {Corbel}, S. 2008, \apj, 688, 1288

\bibitem[{{Piro} \& {Bildsten}(2007)}]{2007ApJ...663.1252P}
{Piro}, A.~L., \& {Bildsten}, L. 2007, \apj, 663, 1252

\bibitem[{{Psaltis} {et~al.}(2000){Psaltis}, {{\"O}zel}, \&
  {DeDeo}}]{2000ApJ...544..390P}
{Psaltis}, D., {{\"O}zel}, F., \& {DeDeo}, S. 2000, \apj, 544, 390

\bibitem[{{Revnivtsev} {et~al.}(2001){Revnivtsev}, {Churazov}, {Gilfanov}, \&
  {Sunyaev}}]{2001A&A...372..138R}
{Revnivtsev}, M., {Churazov}, E., {Gilfanov}, M., \& {Sunyaev}, R. 2001, \aap,
  372, 138

\bibitem[{{Sanna} {et~al.}(2013){Sanna}, {Hiemstra}, {M{\'e}ndez},
  {Altamirano}, {Belloni}, \& {Linares}}]{2013MNRAS.432.1144S}
{Sanna}, A., {Hiemstra}, B., {M{\'e}ndez}, M., {et~al.} 2013, \mnras, 432, 1144

\bibitem[{{Shih} {et~al.}(2005){Shih}, {Bird}, {Charles}, {Cornelisse}, \&
  {Tiramani}}]{2005MNRAS.361..602S}
{Shih}, I.~C., {Bird}, A.~J., {Charles}, P.~A., {Cornelisse}, R., \&
  {Tiramani}, D. 2005, \mnras, 361, 602

\bibitem[{{Strohmayer} \& {Bildsten}(2006)}]{2006csxs.book..113S}
{Strohmayer}, T., \& {Bildsten}, L. 2006, {New views of thermonuclear bursts},
  ed. W.~H.~G. {Lewin} \& M.~{van der Klis}, 113--156

\bibitem[{{Strohmayer} \& {Markwardt}(2002)}]{2002ApJ...577..337S}
{Strohmayer}, T.~E., \& {Markwardt}, C.~B. 2002, \apj, 577, 337

\bibitem[{{Suleimanov} {et~al.}(2012){Suleimanov}, {Poutanen}, \&
  {Werner}}]{2012A&A...545A.120S}
{Suleimanov}, V., {Poutanen}, J., \& {Werner}, K. 2012, \aap, 545, A120

\bibitem[{{Sztajno} {et~al.}(1985){Sztajno}, {van Paradijs}, {Lewin},
  {Trumper}, {Stollman}, {Pietsch}, \& {van der Klis}}]{1985ApJ...299..487S}
{Sztajno}, M., {van Paradijs}, J., {Lewin}, W.~H.~G., {et~al.} 1985, \apj, 299,
  487

\bibitem[{{Taam} {et~al.}(1997){Taam}, {Chen}, \&
  {Swank}}]{1997ApJ...485L..83T}
{Taam}, R.~E., {Chen}, X., \& {Swank}, J.~H. 1997, \apjl, 485, L83

\bibitem[{{van der Klis}(2006)}]{2006csxs.book...39V}
{van der Klis}, M. 2006, {Rapid X-ray Variability}, ed. W.~H.~G. {Lewin} \&
  M.~{van der Klis}, 39--112

\bibitem[{{van Paradijs} {et~al.}(1988){van Paradijs}, {Penninx}, \&
  {Lewin}}]{1988MNRAS.233..437V}
{van Paradijs}, J., {Penninx}, W., \& {Lewin}, W.~H.~G. 1988, \mnras, 233, 437

\bibitem[{{van Paradijs} {et~al.}(1990){van Paradijs}, {van der Klis}, {van
  Amerongen}, {Pedersen}, {Smale}, {Mukai}, {Schoembs}, {Haefner}, {Pfeiffer},
  \& {Lewin}}]{1990A&A...234..181V}
{van Paradijs}, J., {van der Klis}, M., {van Amerongen}, S., {et~al.} 1990,
  \aap, 234, 181

\bibitem[{{Willmore} {et~al.}(1974){Willmore}, {Mason}, {Sanford}, {Hawkins},
  {Murdin}, {Penston}, \& {Penston}}]{1974MNRAS.169....7W}
{Willmore}, A.~P., {Mason}, K.~O., {Sanford}, P.~W., {et~al.} 1974, \mnras,
  169, 7

\bibitem[{{Wilms} {et~al.}(2000){Wilms}, {Allen}, \&
  {McCray}}]{2000ApJ...542..914W}
{Wilms}, J., {Allen}, A., \& {McCray}, R. 2000, \apj, 542, 914

\bibitem[{{Yan} {et~al.}(1998){Yan}, {Sadeghpour}, \&
  {Dalgarno}}]{1998ApJ...496.1044Y}
{Yan}, M., {Sadeghpour}, H.~R., \& {Dalgarno}, A. 1998, \apj, 496, 1044

\bibitem[{{Yoon} {et~al.}(2004){Yoon}, {Langer}, \&
  {Scheithauer}}]{2004A&A...425..217Y}
{Yoon}, S.-C., {Langer}, N., \& {Scheithauer}, S. 2004, \aap, 425, 217

\bibitem[{{Yu} \& {van der Klis}(2002)}]{2002ApJ...567L..67Y}
{Yu}, W., \& {van der Klis}, M. 2002, \apjl, 567, L67

\bibitem[{{Zhang} {et~al.}(2011){Zhang}, {M{\'e}ndez}, \&
  {Altamirano}}]{2011MNRAS.413.1913Z}
{Zhang}, G., {M{\'e}ndez}, M., \& {Altamirano}, D. 2011, \mnras, 413, 1913

\bibitem[{{Zhang} {et~al.}(1995){Zhang}, {Jahoda}, {Swank}, {Morgan}, \&
  {Giles}}]{1995ApJ...449..930Z}
{Zhang}, W., {Jahoda}, K., {Swank}, J.~H., {Morgan}, E.~H., \& {Giles}, A.~B.
  1995, \apj, 449, 930

\bibitem[{{Zhang} {et~al.}(1997){Zhang}, {Lapidus}, {Swank}, {White}, \&
  {Titarchuk}}]{1997IAUC.6541....1Z}
{Zhang}, W., {Lapidus}, I., {Swank}, J.~H., {White}, N.~E., \& {Titarchuk}, L.
  1997, \iaucirc, 6541, 1

\end{thebibliography}
\end{document}